\documentclass[aps,twocolumn,showpacs]{revtex4}
\usepackage{graphicx}
\begin{document}

\title{Snell's Law from an Elementary Particle Viewpoint}
\author{D. Drosdoff and A. Widom}
\affiliation{Physics Department, Northeastern University, Boston MA 02115}

\begin{abstract}
Snell's law of light deflection between media with different indices of
refraction is usually discussed in terms of the Maxwell electromagnetic
wave theory. Snell's law may also be derived from a photon beam theory
of light rays. This latter particle physics view is by far the most simple
one for understanding the laws of refraction.
\end{abstract}
\pacs{42.15.-i,42.15.Dp,41.85.-p}
\maketitle

\section{\label{Intro} Introduction}

Snell's law of refraction is usually discussed in elementary physics
courses wherein the derivations\cite{Becker:1982} depend on the
electromagnetic wave theory of light. The purpose of this note is to show
how the laws of refraction may be derived from the particle (i.e. photon)
view of light rays. In particular, we show how the photon Hamiltonian
\begin{math} H({\bf p},{\bf r}) \end{math} must be derived. In this work,
\begin{math} {\bf p} \end{math} and \begin{math} {\bf r} \end{math}
denote, respectively, the momentum and position of a photon as it
moves along the light ray.

\begin{figure}[bp]
\scalebox {0.8}{\includegraphics{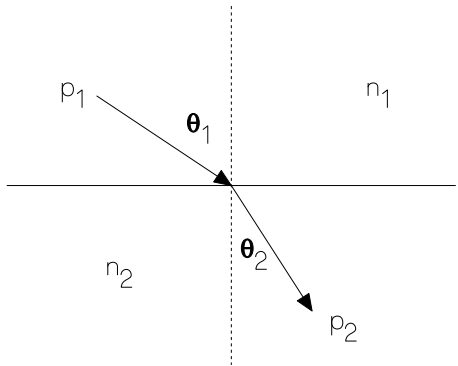}}
\caption{A light ray moves from a medium with index of refraction
$n_1$ into a medium with index of refraction $n_2$. Rays are
considered to be made up of photons with momenta ${\bf p}_1$ and
${\bf p}_2$ respectively.}
\label{sfig1}
\end{figure}

The refraction of a light ray is shown in Fig.\ref{sfig1}. In terms
of the indices of refraction, Snell's law of refraction
asserts\cite{Shirley:1951,Helfgott:2002} that
\begin{equation}
n_1 \sin \theta_1=n_2\sin \theta_2\ \ \ \ {\rm (Snell's\ Law)}.
\label{Intro1}
\end{equation}
The derivation of Eq.(\ref{Intro1}) from the energy and momentum conservation
laws associated with photon deflection is exhibited Sec.\ref{Cons}. The index
of refraction is then {\em defined} in terms of photon energy and momentum.
The relationship between the index of refraction and the photon velocity is
discussed in Sec.\ref{Vel} and from a more general geometric optics limit in
Sec.\ref{Geo}. As an application of the particle viewpoint, we consider
in Sec.\ref{Grav} the gravitational lens, i.e. the bending of light rays
in a gravitational field.

\section{\label{Cons} Conservation Laws}

We assume that the two rays in Fig.\ref{sfig1} are made up of photons
with momenta
\begin{math} {\bf p}_1 \end{math} and \begin{math} {\bf p}_2 \end{math}
respectively. We also assume that the photon energies are
\begin{math} E_1 \end{math} and \begin{math} E_2 \end{math}
respectively. Since there is translational invariance in directions
parallel to the plane separating the two media, the photon momentum
components parallel to the plane are conserved
; i.e.
\begin{equation}
p_1 \sin \theta_1 =p_2 \sin \theta_2.
\label{Cons1}
\end{equation}
The energies of the photons are also conserved; i.e.
\begin{equation}
E_1=E_2.
\label{Cons2}
\end{equation}
In terms of {\em physical} photon energy and momentum, the indices
of refraction are defined by
\begin{equation}
n_1=\frac{cp_1}{E_1}\ \ \ {\rm and}\ \ \ n_2=\frac{cp_2}{E_2}\ .
\label{Cons3}
\end{equation}
Eqs.(\ref{Cons1})-(\ref{Cons3}) imply
\begin{equation}
n_1 \sin \theta_1=n_2\sin \theta_2 \ ,
\label{Cons4}
\end{equation}
which constitutes a simple yet rigorous derivation of Snell's law.

Although similar to other treatments which discuss a {\em formal energy}
and a {\em formal momentum}\cite{Sekiguchi:1987, Dragt:1981}, we
stress that in {\em our} treatment we only refer
to the {\em physical} photon energy and momentum. For example,
\begin{math} E \end{math} has conventional units of {\em Joules}
and momentum \begin{math} {\bf p} \end{math} has conventional units
of {\em Joule-sec/meter}. In other more formal
treatments, the so-called  ``energy'' and ``momentum'' do not have the
usual physical dimensions of energy and momentum.

\section{\label{Vel} Photon Velocity}

The definitions of the indices of refraction in Eq.(\ref{Cons3}) are
equivalent to the more usual definitions
\begin{equation}
n_1=\frac{c}{u_1}\ \ \ {\rm and}\ \ \ n_2=\frac{c}{u_2}\ .
\label{Vel1}
\end{equation}
if one does {\em not in general} identify \begin{math} u \end{math}
with the physical velocity of the photons. The velocity of a photon in
a ray is determined by
\begin{equation}
{\bf v}=\frac{\partial E}{\partial {\bf p}}\ .
\label{Vel2}
\end{equation}
In general for light rays moving through continuous media\cite{Smith:2000}
\begin{equation}
u\ne |{\bf v}|\ \ \ {\rm since}\ \ \
\frac{E}{p}\ne \left|\frac{\partial E}{\partial {\bf p}}\right|.
\label{Vel3}
\end{equation}
In the Maxwell electromagnetic wave
theory\cite{Lifshitz:2000,Sommerfeld:1964, Pokhil:2003} of light,
\begin{math} {\bf u} \end{math} is the {\em phase} velocity
while \begin{math} {\bf v} \end{math} is the {\em group} velocity.
Let us consider this in more detail from a purely particle
physics viewpoint.

\section{\label{Geo} Geometrical Hamiltonian Optics}

In inhomogeneous continuous media, the index of refraction in general
depends on the photon energy as well as position. Eq.(\ref{Cons3})
implies the particle energy restriction
\begin{equation}
E=\frac{c|{\bf p}|}{n({\bf r},E)}\ .
\label{Geo1}
\end{equation}
In principle, the implicit Eq.(\ref{Geo1}) can be solved
for the energy in the form
\begin{equation}
E=H({\bf p},{\bf r}).
\label{Geo2}
\end{equation}
The particle of light, i.e. the photon in the ray, obeys Hamilton's
equations\cite{Hamilton:1931, Landau:2000, Synge:1937, Luneburg:1964};
They are
\begin{equation}
\dot{\bf r}=\frac{\partial H({\bf p},{\bf r})}{\partial {\bf p}}
\ \ \ {\rm and}\ \ \
\dot{\bf p}=-\frac{\partial H({\bf p},{\bf r})}{\partial {\bf r}}\ .
\label{Geo3}
\end{equation}
The velocity of the photon \begin{math} {\bf v}=\dot{\bf r} \end{math}
obeys\cite{Rossi:1965}
\begin{equation}
{\bf v}=\frac{n{\bf u}}
{\left[n+E(\partial n/\partial E)\right]}
\ \ {\rm wherein}\ \ {\bf u}=\frac{E{\bf p}}{p^2}=\frac{c{\bf p}}{np}\ .
\label{Geo4}
\end{equation}
The force on the photon \begin{math} {\bf f}=\dot{\bf p} \end{math}
obeys
\begin{equation}
{\bf f}=\frac{E\ {\bf grad}n}
{\left[n+E(\partial n/\partial E)\right]}\ .
\label{Geo5}
\end{equation}
For a discontinuity in the index of refraction, such as pictured in
Fig.\ref{sfig1}, the impulsive photon force is normal to the surface
of the discontinuity. That there is no force parallel to the surface
of the discontinuity leads directly to Snell's law as in Sec.\ref{Cons}.
We note in passing that Hamilton's Eq.(\ref{Geo3}) follow from
minimizing the optical path length\cite{Born:1970, Newcomb:1983,
Volker:2000}
\begin{equation}
L[{\rm Path},E]=\int_{\rm Path} n({\bf r},E)|d{\bf r}|
\label{Geo6}
\end{equation}
over sufficiently small sections of the path of the ray.

\section{\label{Grav} Photon Deflection due to Gravity}

Astrophysical gravitational lenses are naturally formed and can
be understood on the basis that light rays bend in a gravitational
field. Newton's formulation of gravity attributes the weight of an
object to a gravitational field
\begin{equation}
{\bf g}=-{\bf grad}\Phi .
\label{Grav1a}
\end{equation}
The weight of an object of mass \begin{math} m \end{math} is
\begin{equation}
{\bf w}=m{\bf g}.
\label{Grav1b}
\end{equation}
The gravitational field in a region of space is weak if
for all points within the region the gravitational potential obeys
\begin{equation}
|\Phi ({\bf r})|\ll c^2\ .
\label{Grav2}
\end{equation}

\begin{figure}[tp]
\scalebox {0.6}{\includegraphics{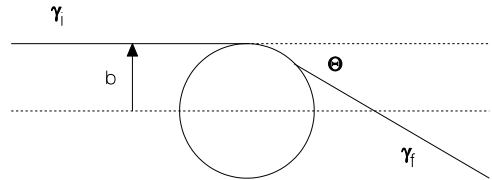}}
\caption{An incident photon denoted by $\gamma_i$
with impact parameter $b$ is refracted from
a spherical gravitational lens. The refraction
index is $n(r)=1+(R_s/r)$. The deflection angle
$\Theta $ describes the final photon denoted by
$\gamma_f$.}
\label{sfig2}
\end{figure}

For a region of space with a weak gravitational field, light rays
bend in a manner described by the index of refraction,
\begin{equation}
n({\bf r})=1-\frac{2\Phi ({\bf r})}{c^2}+\ldots \ ,
\label{Grav3a}
\end{equation}
independently of the energy \begin{math} E \end{math} of the photon.
For example, a spherical astrophysical object will induce in
the neighboring space a potential \begin{math} \Phi=-(GM/r) \end{math}
and thereby a spherical lens with index of refraction
\begin{equation}
n(r)=1+\frac{2GM}{c^2r}+\ldots \ ,
\label{Grav3}
\end{equation}
To a sufficient degree of accuracy, Eqs.(\ref{Geo1}), (\ref{Grav2})
and (\ref{Grav3}) imply
\begin{equation}
|{\bf p}|^2\approx
\left(\frac{E}{c}\right)^2\left(1+\frac{4GM}{c^2r}\right).
\label{Grav4}
\end{equation}
The identities
\begin{eqnarray}
r^2|{\bf p}|^2 &=& |{\bf r\times p}|^2+|{\bf r\cdot p}|^2,
\nonumber \\
r^2|{\bf p}|^2 &=& J^2+r^2p_r^2,
\label{Grav5}
\end{eqnarray}
and
\begin{eqnarray}
J &=& \frac{E b}{c}\ \ \ {(\rm angular\ momentum)},
\nonumber \\
R_s &=& \frac{2GM}{c^2}\ \ \ {(\rm gravitational\ radius)}
\label{Grav6}
\end{eqnarray}
together with Eq.(\ref{Grav4}) imply a radial photon momentum
\begin{equation}
p_r=\pm \frac{E}{c}
\sqrt{1+\left(\frac{2R_s}{r}\right)-\left(\frac{b}{r}\right)^2},
\label{Grav7}
\end{equation}
and its associated scattering angular deflection
\begin{equation}
\Theta=\pi -2\int_{r_{min}}^\infty \frac{bdr}
{r^2\sqrt{1+(2R_s/r)-(b/r)^2}}
\label{Grav8}
\end{equation}
as shown in Fig.(\ref{sfig2}).
The impact parameter thereby obeys
\begin{equation}
b=-R_s\cot (\Theta /2).
\label{Grav9}
\end{equation}
For small angles and large impact parameters\cite{Landau:2000,Einstein:1952},
\begin{equation}
\Theta = -\frac{2R_s}{b}=-\frac{4GM}{c^2b}
\ \ \ \ (b\gg R_s),
\label{Grav10}
\end{equation}
which was used by Einstein to predict the bending of
light around the sun\cite{Lebach:1995, Shapiro:2004}. We note in passing,
the geometrical optics refraction cross section
\begin{equation}
\frac{d\sigma }{d\Omega }
=\left|\frac{bdb}{\sin\Theta d\Theta}\right|
=\frac{R_s^2}{4\sin^4(\Theta /2)}\ .
\label{Grav11}
\end{equation}

\section{\label{Conclude} Conclusions}

The laws of refraction have been shown to follow most easily
by employing a classical Hamiltonian,
\begin{math} E=H({\bf p},{\bf r}) \end{math},
for a photon moving through transparent continuous media.
Snell's law of refraction at the boundary of two
different media is then a simple consequence of the conservation
laws inherent in the Hamiltonian description of the photon.
The use of ray optics in the design of
lens systems is well known. We have here illustrated
the use of the photon Hamiltonian for the case of
gravitational lens effects in astronomy.


\begin{thebibliography}{20}

\bibitem{Becker:1982}
R. Becker, {\it Electromagnetic Fields and Interactions},
page 167 (Dover Publications, Inc., New York, 1982).

\bibitem{Shirley:1951}
J. W. Shirley,
{\it Am. J. Phys.} {\bf 19}, 507 (1951).

\bibitem{Helfgott:2002}
H. Helfgott and M. Helfgott,
{\it Am. J. Phys.} {\bf 70}, 1224 (2002).

\bibitem{Sekiguchi:1987}
T. Sekiguchi and K. Wolf,
{\it Am. J. Phys.} {\bf 55}, 830 (1987).

\bibitem{Dragt:1981}
A. J. Dragt,
{\it Lectures on Nonlinear Orbit Dynamics}
{\bf 87} of American Institute of Physics Conference
Proceedings, (American Institute of Physics, New York, 1981).

\bibitem{Smith:2000}
D. R. Smith and N. Kroll
{\it Physical Review Letters} {\bf 85}, 2933 (2000).

\bibitem{Lifshitz:2000}
L. D. Landau, E. M. Lifshitz, L. P. Pitaevskii,
{\it Electrodynamics of Continuous Media}
(Butterworth-Heinemann, Oxford, 2000), 2nd ed.

\bibitem{Sommerfeld:1964}
A. Sommerfeld,
{\it Optics}
(Academic Press Inc., New York, 1964).

\bibitem{Pokhil:2003}
B. K. Lubsandorzhiev, P. G. Pokhil, R. V. Vasiliev
and Y. E. Vyatchin,
{\it Nucl.Instrum.Meth. A} {\bf 502}, 168 (2003).

\bibitem{Hamilton:1931}
W. R. Hamilton,
{\it The Mathematical Papers of William Rowan Hamilton}
(The University Press, Cambridge, 1931).

\bibitem{Landau:2000}
L. D. Landau and E. M. Lifshitz,
{\it The Classical Theory of Fields}
(Butterworth-Heinemann, Oxford, 2000), 4th ed.

\bibitem{Synge:1937}
J. L. Synge,
{\it Geometrical Optics}
(The University Press, Cambridge, 1937).

\bibitem{Luneburg:1964}
R. K. Luneburg,
{\it Mathematical Theory of Optics}
(University of California Press, Berkeley, 1964).

\bibitem{Rossi:1965}
B. Rossi,
{\it Optics}, page 262
(Addison-Wesley Publishing  Company, Inc.,
Reading, 1965), 3rd ed.

\bibitem{Born:1970}
M. Born and E. Wolf,
{\it Principles of Optics}, page 133
(Pergamon Press, Oxford, 1970), 4th ed.

\bibitem{Newcomb:1983}
W. A. Newcomb,
{\it Am. J. Phys.} {\bf 51}, 338 (1983)

\bibitem{Volker:2000}
P. Volker,
{\it Ray Optics, Fermat's Principle, and Applications
to General Relativity}
(Springer-Verlag, Berlin, 2000).

\bibitem{Einstein:1952}
H. A. Lorentz, A. Einstein, H. Minkowski
and H. Weyl,
{\it The Principle of Relativity}
(Dover Publications, Inc., Toronto, 1952).

\bibitem{Lebach:1995}
D. E. Lebach {\it et al.},
{\it Physical Review Letters} {\bf 75}, 1439 (1995).

\bibitem{Shapiro:2004}
S. S. Shapiro {\it et al.},
{\it Physical Review Letters} {\bf 75}, 121101 (2004).

\end{thebibliography}
\end{document}